\documentclass[
aps,%
11pt,%
final,%
notitlepage,%
oneside,%
onecolumn,%
nobibnotes,%
nofootinbib,%
superscriptaddress,%
noshowpacs,%
centertags]%
{revtex4}
\usepackage{multirow}

\begin{document}

\title{Star-Forming Regions in Dwarf Galaxies  of the Local
Volume}
{\scriptsize Astrophysical Bulletin, Vol. 68, No. 4, 2013
\author{\firstname{S.~S.}~\surname{Kaisin}}
\email{skai@sao.ru}
\author{\firstname{I.~D.}~\surname{Karachentsev}}
\email{ikar@sao.ru}
\affiliation{Special  Astrophysical  Observatory,  Russian  Academy  of  Sciences, Russia}

\begin{abstract}
We present the H$\alpha$ flux measurements for 44 nearby dwarf
galaxies, derived from the observations at the \mbox{6-m} BTA
telescope. H$\alpha$ fluxes were used to determine the rate of
integral star formation of galaxies, SFR. For the observed
galaxies the value of $\log{\rm SFR}$ lies in the range from 0~to
$-8$ $[M_{\odot}$/yr]. The specific star formation rate for all
the sample galaxies does not exceed the limit of  $\log{\rm
SSFR}=-9.2$ [yr$^{-1}$]. A burst of star formation was detected in
the center of a nearby dwarf galaxy UGC\,2172.
\end{abstract}

\maketitle

\section{INTRODUCTION}

Over the past 10 years, the 6-meter BTA telescope of the Special
Astrophysical Observatory has imaged 300 galaxies of the Local
Volume with distances \mbox{$D\sim10$~Mpc} in the Balmer H$\alpha$
line. This number greatly exceeds the total number of H$\alpha$
images of nearby galaxies, acquired at other observatories. The
results of our observations are published in a series of
papers~\cite{kar2005:Kaisin_n,kai2006:Kaisin_n,kar2007:Kaisin_n,kai2008:Kaisin_n,kar2010:Kaisin_n,kai2011:Kaisin_n,kai2013:Kaisin_n}.
Well-exposed H$\alpha$ images of nearby galaxies give an overview
of the structure of star forming regions with the characteristic
linear resolution of the order of 10--30~pc and determine the
integral rate of star formation in the timeline of about 10~Myr. A
comparison of the pattern of star-forming regions, where the young
stars are concentrated, with the distribution of neutral hydrogen
provides an opportunity to determine more accurately the
conditions required for the conversion of gas into stars.

It should be emphasized that about 75\% of the Local Volume
population are dwarf galaxies, where chaotic turbulent motions
dominate over the ordered Keplerian motions typical for massive
spirals. The depth of the potential well in dwarf galaxies is not
large and velocities of about 50~km/s can exceed the parabolic
escape velocity which also imposes an imprint on the features of
star formation in dwarf systems. Because of the shallow depth of
the potential well, many dwarf galaxies easily lose their gas
component, passing through the dense regions of the halo of
massive neighbors. This makes them sensitive indicators of the
dynamical and physical conditions in galaxy groups of different
multiplicity.

The summary of observational data on the star formation rate in
approximately 600 Local Volume galaxies was presented in the
Updated Nearby Galaxy Catalog~\cite{kar2013:Kaisin_n}, and a
collection of H$\alpha$ images of nearby galaxies is contained in
the Local Volume Galaxy Database~\cite{kai2012:Kaisin_n} at the
website {\tt http://www.sao.ru/lv/lvgdb}. In this paper, we
present the H$\alpha$ images and SFR estimates for other 44 Local
Volume galaxies, which (except for three) are dwarf objects with
the absolute magnitudes $M_B$ fainter than $-17^{\rm m}$.

\section{OBSERVATIONS AND DATA REDUCTION}
The images of galaxies in the line of H$\alpha$ and in the nearby
continuum were made from October 2008 to December 2012 with a
typical seeing  of \linebreak \mbox{$1\farcs0$--$2\farcs5$.} The
observations were performed in the primary focus of the 6-m BTA
telescope with the SCORPIO focal reducer~\cite{afa2005:Kaisin_n}
equipped with a CCD chip  $2048\times2048$~px in the $2\times2$
binning mode. With the scale of $0\farcs185$ per pixel, the CCD
provides a field of view sized  \mbox{$6\farcm1 \times 6\farcm1$}.
The images in H$\alpha+[{\rm N\,II}]$ are obtained through a
narrow-band interference filter with bandwidth
$\Delta\lambda=75$~\AA\ and \mbox{$\lambda_{\rm eff}=6555$~\AA}.
For the images in the continuum, we used medium-bandwidth filters
SED\,607 with \mbox{$\Delta\lambda=167$~\AA}, \mbox{$\lambda_{\rm
eff}=6063$~\AA} \ and SED\,707 with \linebreak
\mbox{$\Delta\lambda=707$~\AA}, \mbox{$\lambda_{\rm
eff}=7036$~\AA}. Typical exposure time was $2\times600$~s in
H$\alpha$ and $2\times300$~s in the continuum. Given a small range
of line-of-sight velocities of galaxies, $V\leq600$ km/s, we
managed with the same H$\alpha$ filter.

We used a standard procedure of data reduction. The original
images had bias subtracted and then flat-field corrected. After
the removal of traces of cosmic ray particles and sky background
subtraction, the images for each object were combined. Finally,
all the images in the continuum were normalized to the
\mbox{H$\alpha$} image using 7--20 field stars and then
subtracted. From the continuum-subtracted H$\alpha$ images, we
have measured the integral \mbox{H$\alpha$} fluxes of galaxies
using the images of the spectrophotometric standard
stars~\cite{oke1990:Kaisin_n}, which were taken at the same nights
as the objects. The formal measurement accuracy of  the integral
fluxes  was about 10\%.

\section{RESULTS OF OBSERVATIONS}
The Appendix shows a mosaic of images of 44~galaxies we have
observed. The left-hand side images in each pair represent the
total exposure in the H$\alpha$ line and in the continuum, and the
right-hand side images correspond to the
``H$\alpha$\,--\,continuum'' difference. The lower corners of
right-hand side images specify the angular scale and orientation
``\mbox{North--East}.''

After the subtraction of the continuum, many images in the line of
H$\alpha$ reveal the ``stubs'' of stars, caused by the difference
in seeing, as well as the saturation effect in  bright stars or
abnormal color indices in some stars. This limits the accuracy of
finding the integral H$\alpha$ flux of galaxies, especially in the
objects of low surface brightness or in the galaxies located at
low galactic latitudes, where the Galactic background stars
abound.

For each galaxy represented in the mosaic (see the Appendix), we
identified the integral flux in the line of H$\alpha$ or its upper
limit in the units of erg/(cm$^2$\,s). The observed $F_{\rm
H\alpha}$ flux, corrected for the extinction of light in the
Galaxy according to~\cite{schle1998:Kaisin_n}, was used to
estimate the integral SFR, following the relation of
Kennicutt~\cite{ken1998:Kaisin_n}
$$
\log{\rm SFR}=\log F_{\rm H\alpha}+2\log D+8.98.
$$
Here $D$ is the distance to the galaxy in Mpc, and the SFR value
is expressed in units of  $M_{\odot}$/yr.

We neglected the internal absorption in the dwarf galaxy itself as
well as  the contribution from the nearby H$\alpha$ emission
doublet [N\,II], since both of these effects are small for the
galaxies of low
luminosities~\cite{ver2001:Kaisin_n,lee2009:Kaisin_n}.

An overview of data on the galaxies we observed is presented in
the table. Its columns contain: (1)~the name of the galaxy;
(2)~equatorial coordinates for the epoch J\,2000.0;
\mbox{(3)--(5)} integral apparent magnitude, morphological type,
and the distance (Mpc), according to the UNGC
catalog~\cite{kar2013:Kaisin_n}; (6),~(7)~the logarithm of the
observed flux in the lines of H$\alpha + [{\rm N\,II}]$ and its
measurement error; (8)~the logarithm of the integral SFR;
\mbox{(9),~(10)}~dimensionless parameters \mbox{${\rm P}=\log({\rm
SFR}\times T_0/M_*)$} and \mbox{${\rm F}=\log(1.85\,M_{\rm
H\,I}/{\rm SFR}\times T_0$)}, which characterise the evolutionary
status of the galaxy, having a stellar mass $M_*$ and hydrogen
mass $M_{\rm H\,I}$, at the cosmic timescale of $T_0=13.7\times
10^9$ yrs; the values of $M_*$ and $M_{\rm H\,I}$ are adopted from
the UNGC catalog~\cite{kar2013:Kaisin_n}. The three last columns
of the table list the values of $\log F_{\rm H\alpha}$ and $\log
{\rm SFR}_{\rm H\alpha}$ from the summary
of~\cite{ken2008:Kaisin_n}, determined by other authors. For
comparison, the last column of our table shows the SFR estimates
obtained in~\cite{Kar2013:Kaisin_n} from the ultraviolet flux
(FUV) measured by the GALEX space
telescope~\cite{gil2007:Kaisin_n}.

\begin{turnpage}
\begin{table*}[]
\setcaptionmargin{0mm} \onelinecaptionstrue
\captionstyle{nonumber} \caption{\centerline{Parameters of  44
nearby dwarf galaxies}}
\medskip
\begin{tabular}{l|c|c|r|r|r|c|r|r|r|c|c|r}
 \hline
Name & J\,2000.0 & $B_t$ & \multicolumn{1}{c|}{$T$} & \multicolumn{1}{c|}{$D$} & \multicolumn{1}{c|}{$\log F_{\rm obs}$} & Err & \multicolumn{1}{c|}{$\log {\rm SFR}$} & \multicolumn{1}{c|}{$\rm P$} & \multicolumn{1}{c|}{$\rm F$} & $\log F_{\rm lit}$ & $\log {\rm SFR}_{\rm H\alpha}$ & \multicolumn{1}{c}{$\log {\rm SFR}_{\rm FUV}$} \\
\hline
 (1) & (2) & (3) & \multicolumn{1}{c|}{(4)} & \multicolumn{1}{c|}{(5)} & \multicolumn{1}{c|}{(6)} & (7) & \multicolumn{1}{c|}{(8)} & \multicolumn{1}{c|}{(9)} & \multicolumn{1}{c|}{(10)} & (11) & (12) & \multicolumn{1}{c}{(13)}\\
\hline
 UGC\,12894       &    000022.5+392944   & 16.80 &  10  &  8.5 & $-13.23$  & $\pm 0.01$ & $-2.29$  &  0.27    &  0.35   & $-13.42\pm 0.03$  & $-2.48$  &  $-2.03$\\
 AGC\,748778      &    000634.4+153039   & 18.90 &  10  &  5.4 & $-15.27$  & $\pm 0.24$ & $-4.76$  & $-0.90$  &  1.41   &   $-$            & $-$       &  $-3.65$\\
 UGC\,00064       &    000744.0+405232   & 15.50 &  10  &  9.6 & $-12.52$  & $\pm 0.01$ & $-1.51$  &  0.47    &  0.22   &   $-$            & $-$       &  $-1.63$\\
 UGC\,1561        &    020405.0+241228   & 14.51 &   9  &  7.2 & $-12.88$  & $\pm 0.01$ & $-1.89$  & $-0.28$  & $-0.13$ & $-12.90\pm 0.04$  & $-1.91$  &  $-1.89$\\
 DDO\,019         &    022500.2+360216   & 15.80 &  10  &  9.3 & $-12.83$  & $\pm 0.01$ & $-1.84$  &  0.28    &  0.30   & $-12.73\pm 0.06$  & $-1.74$  &  $-1.67$\\
 Halogas          &    022720.0+335730   & 18.00 &  10  &  9.3 & $-13.80$  & $\pm 0.02$ & $-2.81$  &  0.20    &  0.07   &   $-$            & $-$       &  $-3.02$\\
 DDO\,025         &    023318.2+332928   & 13.96 &   8  &  9.3 & $-12.69$  & $\pm 0.01$ & $-1.69$  & $-0.42$  &  0.23   & $-12.41\pm 0.05$  & $-1.40$  &  $-1.33$\\
 DDO\,024         &    023343.0+403141   & 13.68 &   8  &  9.8 & $-12.59$  & $\pm 0.01$ & $-1.58$  & $-0.40$  &  0.44   & $-12.51\pm 0.11$  & $-1.50$  &  $-$\\
 UGC\,02172       &    024210.8+432119   & 14.60 &  10  &  9.3 & $-11.95$  & $\pm 0.01$ & $-0.94$  &  0.88    & $-1.05$ &   $-$            & $-$       &  $-$\\
 KKH\,22          &    034456.6+720352   & 18.00 &  10  &  3.5 & $<-15.25$ & $\pm 0.14$ & $<-4.82$ & $<-1.49$ & $>1.42$ &   $-$            & $-$       &  $<-4.05$\\
 UGC\,03501       &    063838.4+491530   & 16.70 &  10  & 15.5 & $-13.73$  & $\pm 0.02$ & $-2.26$  & $-0.28$  &  0.71   &   $-$            & $-$       &  $-1.65$\\
 KKH\,38          &    064754.9+473050   & 17.40 &  10  & 19.3 & $-13.58$  & $\pm 0.23$ & $-1.95$  &  0.16    &  0.82   &   $-$            & $-$       &  $-$\\
 HIZSS\,003\,B    &    $070024.7-041318$ & 18.00 &  10  &  1.6 & $-13.52$  & $\pm 0.01$ & $-3.09$  & $-0.32$  &  0.48   & $-13.66\pm 0.08$  & $-3.23$  &  $-$\\
 HIZSS\,003\,A    &    $070029.3-041230$ & 19.00 &  10  &  1.6 & $-14.60$  & $\pm 0.09$ & $-4.17$  & $-1.40$  &  1.55   &   $-$            & $-$       &  $-$\\
 AGC\,174585      &    073610.3+095911   & 17.90 &  10  &  6.1 & $-14.17$  & $\pm 0.03$ & $-3.58$  & $-0.18$  &  0.40   &   $-$            & $-$       &  $-$\\
 KKH\,40          &    074656.4+511746   & 16.60 &  10  &  7.0 & $-13.38$  & $\pm 0.02$ & $-2.66$  &  0.07    &  0.27   &   $-$            & $-$       &  $-2.47$\\
 AGC\,174605      &    075021.7+074740   & 18.00 &  10  &  6.0 & $-14.53$  & $\pm 0.05$ & $-3.97$  & $-0.49$  &  0.86   &   $-$            & $-$       &  $<-4.79$\\
 NGC\,2541        &    081440.1+490342   & 12.26 &   7  & 12.4 & $-11.56$  & $\pm 0.01$ & $-0.28$  &  0.37    & $-0.06$ & $-11.68\pm 0.02$  & $-0.41$  &  $+0.09$\\
 UMa\,II          &    085130.0+630748   & 14.80 & $-2$ &  0.0 & $<-15.23$ & $\pm 0.14$ & $<-9.16$ & $<-3.22$ & $>2.10$ &   $-$            & $-$       &  $-8.84$\\
 UGC\,04787       &    090734.9+331636   & 14.60 &   8  & 20.3 & $-12.86$  & $\pm 0.01$ & $-1.19$  & $-0.32$  &  0.16   & $-12.82\pm 0.07$  & $-1.15$  &  $-0.82$\\
 LV J\,0913+1937  &    091339.0+193708   & 17.40 &  10  &  4.4 & $-13.58$  & $\pm 0.04$ & $-3.26$  &  0.21    & $-0.14$ &   $-$            & $-$       &  $-3.36$\\
 UGC\,04879       &    091602.2+525024   & 13.80 &   9  &  1.3 & $-13.60$  & $\pm 0.03$ & $-4.34$  & $-1.19$  &  0.46   & $-13.70\pm 0.18$  & $-4.44$  &  $-3.29$\\
 \hline
\end{tabular}
\end{table*}
\end{turnpage}

\begin{turnpage}
\setcounter{table}{0}
\begin{table*}[]
\setcaptionmargin{0mm} \onelinecaptionstrue
\captionstyle{nonumber} \caption{\centerline{Parameters of  44
nearby dwarf galaxies. (Contd.)}}
\medskip
\begin{tabular}{l|c|c|r|r|r|c|r|r|r|c|c|r}
 \hline
Name & J\,2000.0 & $B_t$ & \multicolumn{1}{c|}{$T$} & \multicolumn{1}{c|}{$D$} & \multicolumn{1}{c|}{$\log F_{\rm obs}$} & Err & \multicolumn{1}{c|}{$\log {\rm SFR}$} & \multicolumn{1}{c|}{$\rm P$} & \multicolumn{1}{c|}{$\rm F$} & $\log F_{\rm lit}$ & $\log {\rm SFR}_{\rm H\alpha}$ & \multicolumn{1}{c}{$\log {\rm SFR}_{\rm FUV}$} \\
\hline
 (1) & (2) & (3) & \multicolumn{1}{c|}{(4)} & \multicolumn{1}{c|}{(5)} & \multicolumn{1}{c|}{(6)} & (7) & \multicolumn{1}{c|}{(8)} & \multicolumn{1}{c|}{(9)} & \multicolumn{1}{c|}{(10)} & (11) & (12) & \multicolumn{1}{c}{(13)}\\
\hline
 UGC\,04932       &    091934.1+510633   & 15.17 &   8  & 20.6 & $-13.26$  & $\pm 0.03$ & $-1.61$  & $-0.16$  &  0.47   &   $-$            & $-$       &  $-1.26$\\
 UGC\,04998        &   092512.1+682259   & 15.00  &   9  &  8.2 & $-13.61$  & $\pm0.02$ & $-2.74$  & $-1.31$  & $-0.09$ & $-13.27\pm 0.09$  & $-2.40$   &  $-2.55$\\
 NGC\,2903-H\,I-1  &   093039.9+214325   & 18.20  &  10  &  8.9 & $-13.84$  & $\pm0.02$ & $-2.93$  &  0.28    & $-0.51$ &   $-$            & $-$       &  $-3.54$\\
 LV J\,1018+4109   &   101822.2+410957   & 18.40  & $-1$ & 11.1 & $<-15.33$ & $\pm0.26$ & $<-4.25$ & $<-1.82$ &         &   $-$            & $-$       &  $<-4.28$\\
 NGC\,3239         &   102504.9+170949   & 11.73  &   8  &  7.9 & $-11.29$  & $\pm0.01$ & $-0.45$  &  0.17    & $-0.53$ & $-11.32\pm 0.03$  & $-0.47$   &  $-0.40$\\
 LeG\,06           &   103955.7+135428   & 18.30  &  10  & 10.4 & $<-15.36$ & $\pm0.26$ & $<-4.31$ & $<-1.21$ & $>1.30$ &   $-$            & $-$       &  $-3.55$\\
 LeG\,19           &   104654.8+124717   & 17.80  & $-1$ & 10.4 & $<-15.34$ & $\pm0.28$ & $<-4.30$ & $<-2.08$ &         &   $-$            & $-$       &  $<-4.43$\\
 KDG\,078          &   112954.0+522414   & 16.70  &  10  &  8.8 & $<-15.35$ & $\pm0.25$ & $<-4.45$ & $<-1.82$ & $>1.33$ &   $-$            & $-$       &  $<-4.46$\\
 LV J\,1217+4703   &   121710.1+470349   & 18.50  &  10  &  7.8 & $<-15.37$ & $\pm0.21$ & $<-4.59$ & $<-1.1$2 & $>1.16$ &   $-$            & $-$       &  $-4.89$\\
 KK\,138           &   122158.4+281434   & 18.70  &  10  &  6.3 & $<-15.26$ & $\pm0.23$ & $-2.82$  & $-0.10$  &  0.15   &   $-$            & $-$       &  $-2.67$\\
 LV J\,1228+4358   &   122844.9+435818   & 14.20  &  10  &  4.0 & $<-15.22$ & $\pm0.23$ & $<-5.01$ & $<-2.70$ &         &   $-$            & $-$       &  $<-5.13$\\
 KK\,152           &   123324.9+332105   & 16.30  &  10  &  6.9 & $-13.70$  & $\pm0.02$ & $-3.02$  & $-0.33$  &  0.70   &   $-$            & $-$       &  $-2.48$\\
 UGCA\,292         &   123840.0+324560   & 16.07  &  10  &  3.6 & $-12.65$  & $\pm0.01$ & $-2.54$  &  0.86    &  0.11   & $-12.76\pm 0.01$  & $-2.65$   &  $-2.59$\\
 BTS\,146          &   124002.1+380002   & 17.50  &  10  &  8.5 & $-15.47$  & $\pm0.11$ & $-4.6$2  & $-1.63$  &  1.72   &   $-$            & $-$       &  $-3.44$\\
 KDG\,192          &   124345.0+535732   & 16.60  &  10  &  7.4 & $-13.38$  & $\pm0.04$ & $-2.65$  &  0.11    &  0.67   &   $-$            & $-$       &  $-2.42$\\
 LV J\,1243+4127   &   124355.7+412725   & 17.20  &  10  &  6.1 & $-15.12$  & $\pm0.07$ & $-4.55$  & $-1.39$  &  1.71   &   $-$            & $-$       &  $-3.33$\\
 KK\,191           &   131339.7+420239   & 18.20  &  10  &  6.0 & $<-15.29$ & $\pm0.24$ & $<-4.74$ & $<-1.16$ & $>1.08$ &   $-$            & $-$       &  $-4.96$\\
 KDG\,235          &   170025.3+701724   & 16.80  &  10  & 10.6 & $-13.80$  & $\pm0.06$ & $-2.74$  & $-0.25$  &  0.74   & $-14.45\pm 0.19$  & $-3.39$   &  $-2.48$\\
 ALFA~ZOA          &   195211.8+142824   & 16.90  &   9  &  7.1 & $-14.00$  & $\pm0.02$ & $-3.08$  & $-0.59$  &  0.27   &   $-$            & $-$       &  $-$\\
 KK\,258           &   $224043.9-304759$ & 16.30  & $-3$ &  2.0 & $-14.37$  & $\pm0.08$ & $-4.78$  & $-1.70$  &  0.57   &   $-$            & $-$       &  $-4.58$\\
 Pisces\,II        &   225831.0+055709   & 17.20  & $-3$ &  0.1 & $-15.01$  & $\pm0.07$ & $-7.45$  & $-2.02$  &         &   $-$            & $-$       &  $-7.56$ \\
\hline
\end{tabular}
\end{table*}
\end{turnpage}

\section{FEATURES OF SOME OBSERVED GALAXIES}
As we can see from the table, about 3/4 of this sample are
irregular dwarf galaxies of \linebreak Ir, Im $(T=10, 9)$  types
and blue compact  BCD galaxies  $(T = 9)$. The remaining quarter
of the sample accounts for late-type dwarf spirals Sdm, \linebreak
Sm $(T = 7, 8)$ and dwarf spheroidal systems  $(T<0)$. For the
irregular and BCD galaxies the presence of one or several emission
knots is typical. In some cases, compact H\,II-regions are
immersed in a diffuse emission environment of various contrast.
Some irregular galaxies of low surface brightness (e.g., KKH\,22,
KDG\,78) do not exhibit notable H$\alpha$ fluxes. Let us describe
some of the most interesting objects in the studied sample.

{\it UGC\,2172}. This irregular galaxy with an absolute magnitude
of $M_B=-15.69$ is in its starburst activity phase. A bright star
is projected to the north-west from its center. The major emission
from UGC\,2172 is concentrated in its central area, from which the
low-contrast arcuate filaments are stretching to the periphery.
According to the structure of the emission arcs, this galaxy is
similar to a nearby example of a starburst galaxy, NGC\,1569.

{\it HIZSS\,03\,A+B}. A tight pair of irregular galaxies with the
distance between the centers of $1\farcm4$, or 0.7~kpc. It lies
almost exactly in the plane of the Galaxy at the latitude of
$b=-0\fdg1$. Despite the strong extinction, Silva et
al.~\cite{sil2005:Kaisin_n} have determined the distance to the
pair of 1.67~Mpc with the tip of the red giant branch method.
According to the observations in the line of H\,I
21~cm~\cite{beg2005:Kaisin_n}, the components of the pair are well
resolved kinematically and have the radial velocity difference of
35~km/s. In the western, more compact component, the emission in
H$\alpha$ was for the first time discovered by Massey et
al.~\cite{mas2003:Kaisin_n}.   Apparently, this pair of dwarf
galaxies is the closest representative of a binary system in the
phase immediately before the merger.

{\it NGC\,2541}. This late-type (Sdm) spiral galaxy has the
absolute magnitude of $M_B=-18.71$ which makes it the brightest in
the considered sample. The H$\alpha$ image reveals an ensemble of
compact star-forming regions organized in a flocculent spiral
structure.

{\it UMa\,II}. A spheroidal dwarf companion of our Galaxy of
extremely low surface brightness, recently discovered by stellar
counts~\cite{zuk2006:Kaisin_n}. Having an angular diameter  of
about $25^{\prime}$, our H$\alpha$ image of the object could only
cover the central region of UMa\,II.

{\it NGC\,2903-H\,I-1}. A blue compact satellite of a giant spiral
galaxy NGC\,2903, discovered within the extragalactic H\,I Arecibo
Legacy Fast ALFA (\mbox{ALFALFA}) survey~\cite{irw2009:Kaisin_n}.
With the absolute magnitude of $M_B=-11.68$ it can be considered
an intergalactic H\,II-region on the distant outskirts of the disk
of NGC\,2903.

{\it NGC\,3239 $=$ Arp\,263 $=$ VV\,095}. An interacting pair of
irregular galaxies with two curved tails and powerful star-forming
regions. Active state of this exotic system is obviously caused by
the ongoing processes of dynamic merging of its components.

{\it LV~J\,1228+4358}.  A dwarf spheroidal galaxy possessing a
very low surface brightness, with structure distorted by the tidal
influence of the NGC\,4449 galaxy. Found  by Karachentsev et
al.~\cite{Kar2007:Kaisin_n} and studied in detail by
Martinez-Delgado~\cite{mar2012:Kaisin_n}.

{\it UGCA\,292}. A ragged blue galaxy of low luminosity
($M_B=-11.79)$, looming in the shape of an arc over a bright star.
UGCA\,292 contains a large amount of neutral hydrogen and is
catalogued as one of the lowest metallicity objects in the CVn\,I
cloud~\cite{van2000:Kaisin_n}.

{\it ALFA~ZOA~J\,1952+1428}. A compact blue galaxy in the Zone of
Avoidance (ZOA) of the Milky Way, detected in the blind H\,I
survey at Arecibo~\cite{mci2011:Kaisin_n}. Located near the center
of the Local Tully Void, it is an extremely isolated object of the
Local Volume.

{\it KK\,258 $=$ ESO\,468-020}. An isolated dwarf galaxy of
intermediate type between dIr and dSph. Our image, taken not far
above the technical horizon of the \mbox{6-m} BTA telescope
($\mbox{Dec}\simeq-31^{\circ}$), reveals one compact
\mbox{H$\alpha$ emission} near the center of the galaxy.

{\it Pisces\,II}. A dwarf $(M_B=-4.4)$ spheroidal satellite of our
Galaxy, discovered by Belokurov et al.~\cite{bel2010:Kaisin_n}. In
the optical contour of this dwarf, there is a possible emission
point source, which is likely to be a red star.

\section{DISCUSSION AND CONCLUSIONS}

\begin{figure}[]
\
\includegraphics[scale=0.8]{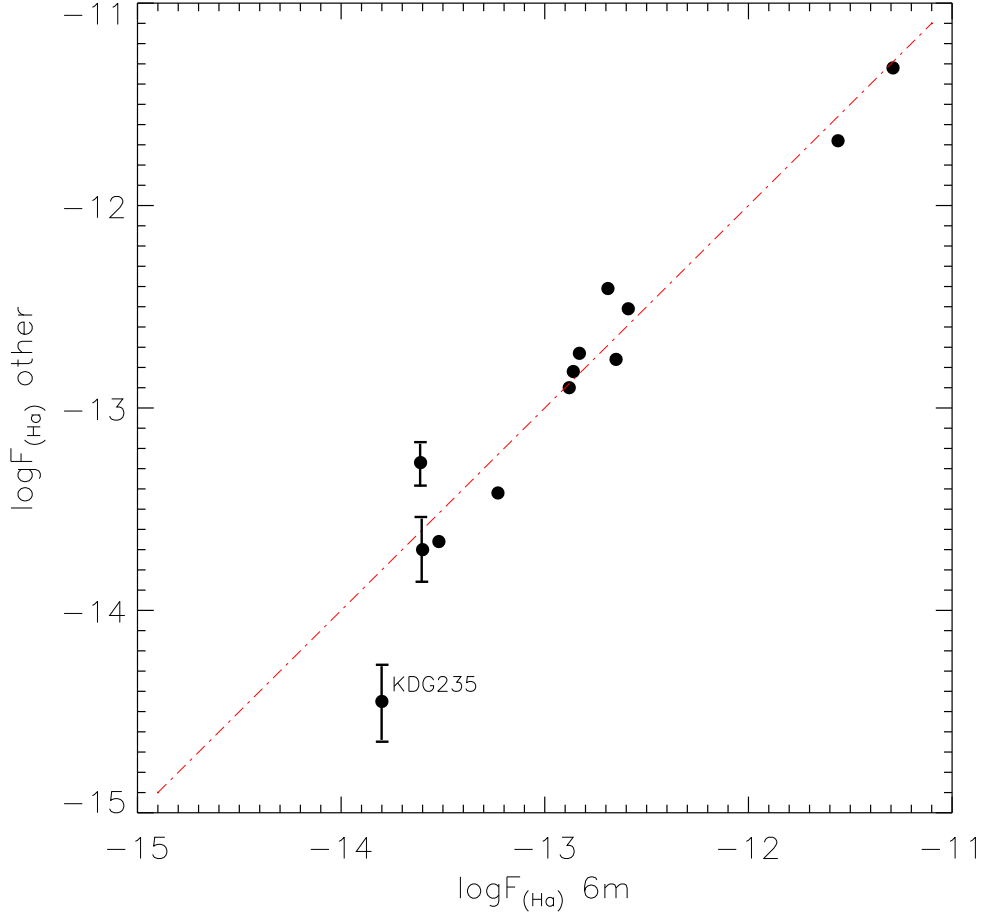}
\captionstyle{normal} \vspace{1mm} \caption{Comparison of integral
H$\alpha$ flux estimates with observations at the 6-m telescope
and data of other authors.} \vspace{3mm}
\end{figure}

Among the 44 galaxies we observed,  $F_{\rm H\alpha}$ fluxes for
13 objects were also measured by other authors. Figure~1 shows the
relationship between our $F_{\rm H\alpha}$ estimates  and
literature data. The figure shows that the scatter of values
relative to the line of \mbox{$\log F_{\rm
H\alpha}(\mbox{6-m})=\log F_{\rm H\alpha}(\mbox{others})$}
slightly increases with decreasing flux. If we exclude the low
surface brightness galaxy KDG\,235, for which the H$\alpha$ flux
in~\cite{ken2008:Kaisin_n} is measured with low accuracy, then
the mean difference of $\log F_{\rm H\alpha}(\mbox{6-m})-\log
F_{\rm H\alpha}(\mbox{others})$ will amount to
\mbox{$-0.01\pm0.05$}, and the standard deviation of the
difference will be equal to 0.16. The last value is two times
greater than the RMS sum of individual flux measurement errors
(0.08). It is clear that the hard-to-control transparency
variations during the observations, as well as the differences
between the methods applied by various authors when taking into
account the diffuse components of H$\alpha$ emission, lead to the
twofold difference between the external and internal flux
measurement errors.

As follows from the data of the last column of the table, most of
the galaxies we observed have their SFRs estimated from the UV
fluxes measured by the Galaxy Evolution Explorer satellite
(GALEX). A comparison of independent values of $\log{\rm SFR}$ is
shown in Fig.~2, where the solid circles correspond to our
measurements of the H$\alpha$ flux, the crosses---to the H$\alpha$
data of other authors, and open circles represent the upper limit
for the observed star formation rate. For the majority of dwarf
galaxies, the H$\alpha$ flux underestimates the SFR value as
compared to the FUV flux. This well-known fact was discussed in
detail by various authors, in particular
in~\cite{lee2009:Kaisin_n} and~\cite{Kar2013:Kaisin_n}. According
to Pflamm-Altenburg et al.~\cite{pfl2007:Kaisin_n}, conditions of
formation of the most massive stars in dwarf and normal spiral
galaxies are somewhat different. An empirical normalization \mbox
{${\rm SFR}_{\rm H\alpha} \simeq {\rm SFR_{\rm FUV}}$}, which has
been made for the spirals, is not fulfilled for the dwarf systems,
and at $\log{\rm SFR}\sim -5$ the scatter of estimates can be more
than one order. This feature is also visible in Fig.~2. Note,
however, that cases exist (for instance, a blue compact galaxy
NGC\,2903-H\,I-1) when the SFR estimate from the \mbox{H$\alpha$
flux} proved to be greater than that from the FUV flux. It is
intriguing that for the extremely faint spheroidal companions of
the Milky Way like UMa\,II and Pisces\,II, the upper levels of SFR
from H$\alpha$ and FUV fluxes turned out to be close to each other
at $\log{\rm SFR}\sim-8$. A similar situation is already noted for
the low-mass companions of M\,31 and
M\,81~\cite{kai2013:Kaisin_n}. In the case of Pisces\,II, two
faint FUV sources and one H$\alpha$ source fall into the optical
path of the galaxy. However, they do not coincide with each other
by the coordinates, probably being the artifacts (background stars
with unusual energy distribution).

\begin{figure}[]

\includegraphics[scale=0.8]{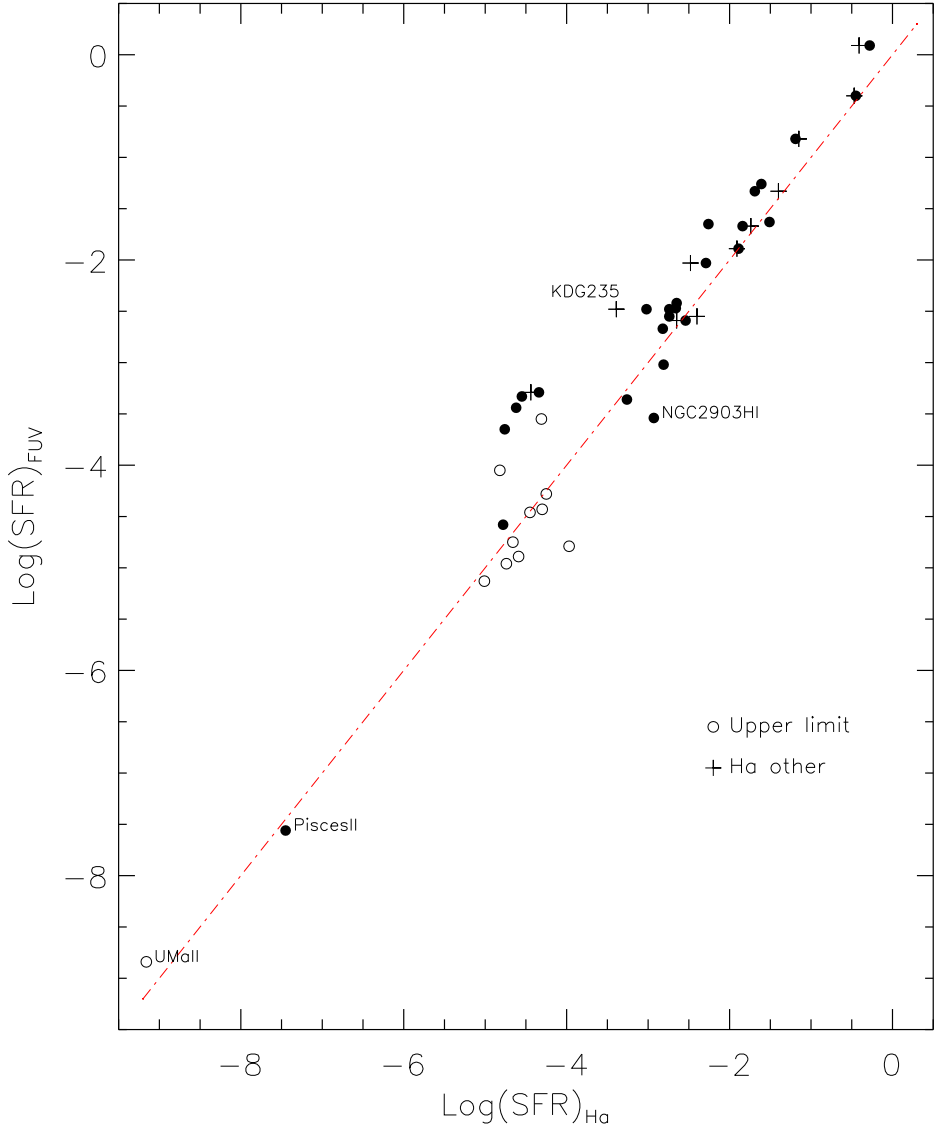}
\captionstyle{normal} \vspace{1mm} \caption{Comparison of the
integral SFR estimates in galaxies obtained from the H$\alpha$
flux and the flux in the far ultraviolet (FUV). Galaxies with an
estimate of the upper limit of SFR are marked with open circles.
The crosses denote the SFR values from the H$\alpha$ flux measured
by other authors.} \vspace{3mm}
\end{figure}

\begin{figure}[]
\includegraphics[scale=0.8]{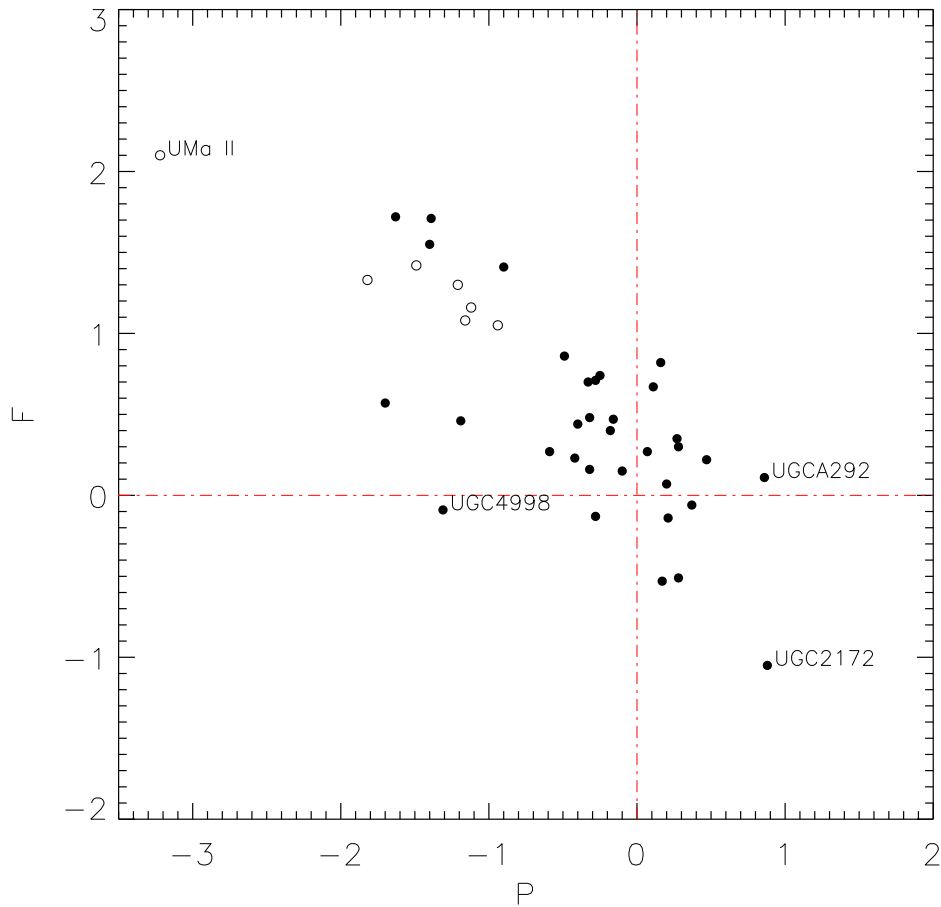}
\captionstyle{normal} \vspace{1mm} \caption{The diagnostic
``Past--Future'' diagram for the observed galaxies. The galaxies
with the upper SFR limit are shown by open circles.} \vspace{3mm}
\end{figure}

As follows from the diagnostic diagram \linebreak ``Past--Future''
(Fig.~3), the majority of objects in our sample are located near
the origin \{${\rm P}=0, {\rm F}=0$\}. This means that at the
observed SFR, the galaxy has time to reproduce its stellar mass on
the cosmological scale $T_0$, while the gas reserves in it are
sufficient to maintain the observed SFR for still another Hubble
time $T_0$. However, there are several galaxies significantly
deviating from the majority. As we have noted above, the UGC\,2172
galaxy is currently in the state of starburst activity. Its
observed SFR is now almost an order of magnitude higher than the
average for the given mass, and its gas reserves will be exhausted
in time of only about $T_0/10$. In the case of a metal-poor dIr
galaxy UGCA\,292, SFR is also very high, but the gas reserves are
sufficient to maintain the observed SFR on the entire Hubble
timescale. The BCD galaxy UGC\,4998 reveals tiny faint
star-forming regions in its central part. In the past, the average
star formation rate in UGC\,4998 was an order of magnitude more
intense than that currently observed.

Analyzing the sample of 627 galaxies of the Local Volume with SFR
estimates both from the H$\alpha$ and FUV fluxes, we noted
in~\cite{Kar2013:Kaisin_n} that the specific star formation rate
(SSFR) per stellar mass unit ${\rm SSFR}=\dot M_*/M_*$   does not
exceed the upper limit of  $\log{\rm SSFR}_{\rm max}\simeq
-9.4$~[yr$^{-1}$] in 99\% of the sample objects. Among the 44
galaxies we have considered, there are only two of the most
active, UGC\,2172 and UGCA\,292, which slightly exceed the said
limit, having their $\log{\rm SSFR}$ of \mbox{$-9.26$} and
$-9.27$~[yr$^{-1}$] respectively. However, the error in
determining the stellar mass of such faint galaxies by their
luminosity can reach up to 50\%. The presence of the maximum
(quasi-Eddington) limit for the SSFR is an important parameter
that characterizes the process of conversion of gas into stars
during the present epoch.

\begin{acknowledgements}
This work was supported by the grant of  the Russian Foundation
for Basic Research  (project \mbox{no.~13-02-92960-IND-a, 13-02-00780-a}) and by
the Ministry of Education and Science of the Russian Federation
(agreement  8523, state contracts no.~14.518.11.7070,
16.518.11.7073).
\end{acknowledgements}

%\end{document}
\onecolumngrid

\section*{APPENDIX}

\includegraphics[scale=0.8]{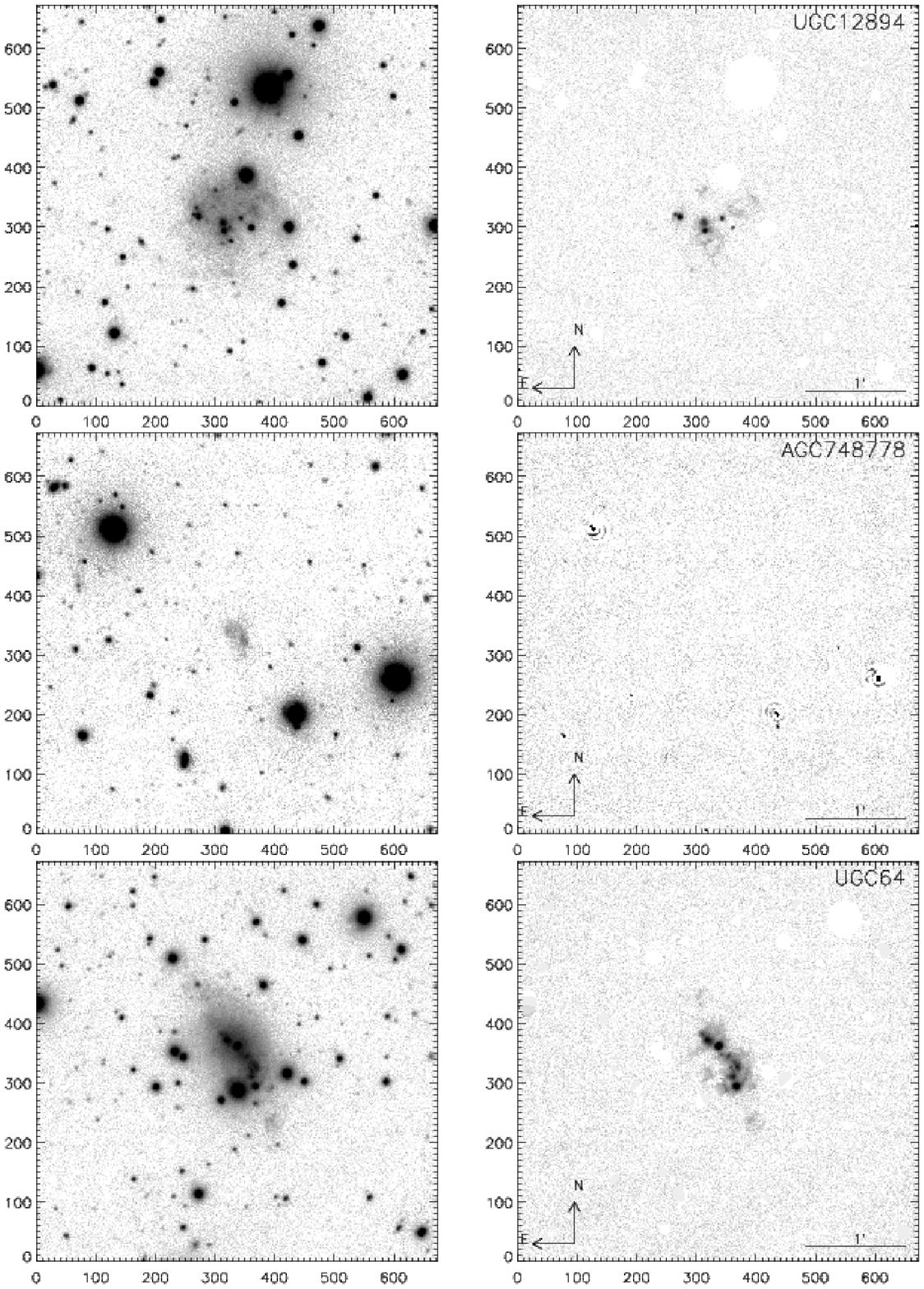}

\includegraphics[scale=0.8]{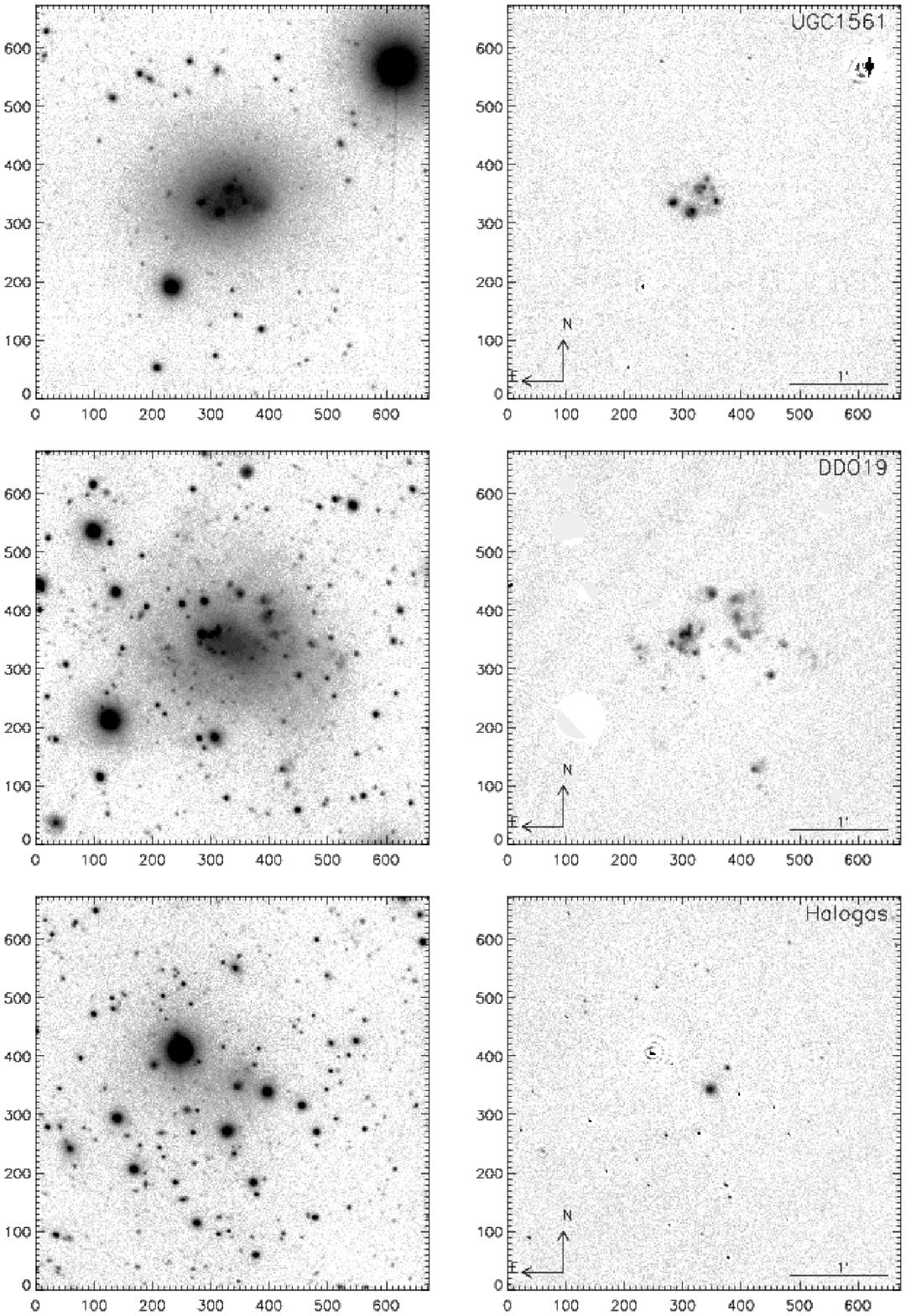}

\includegraphics[scale=0.8]{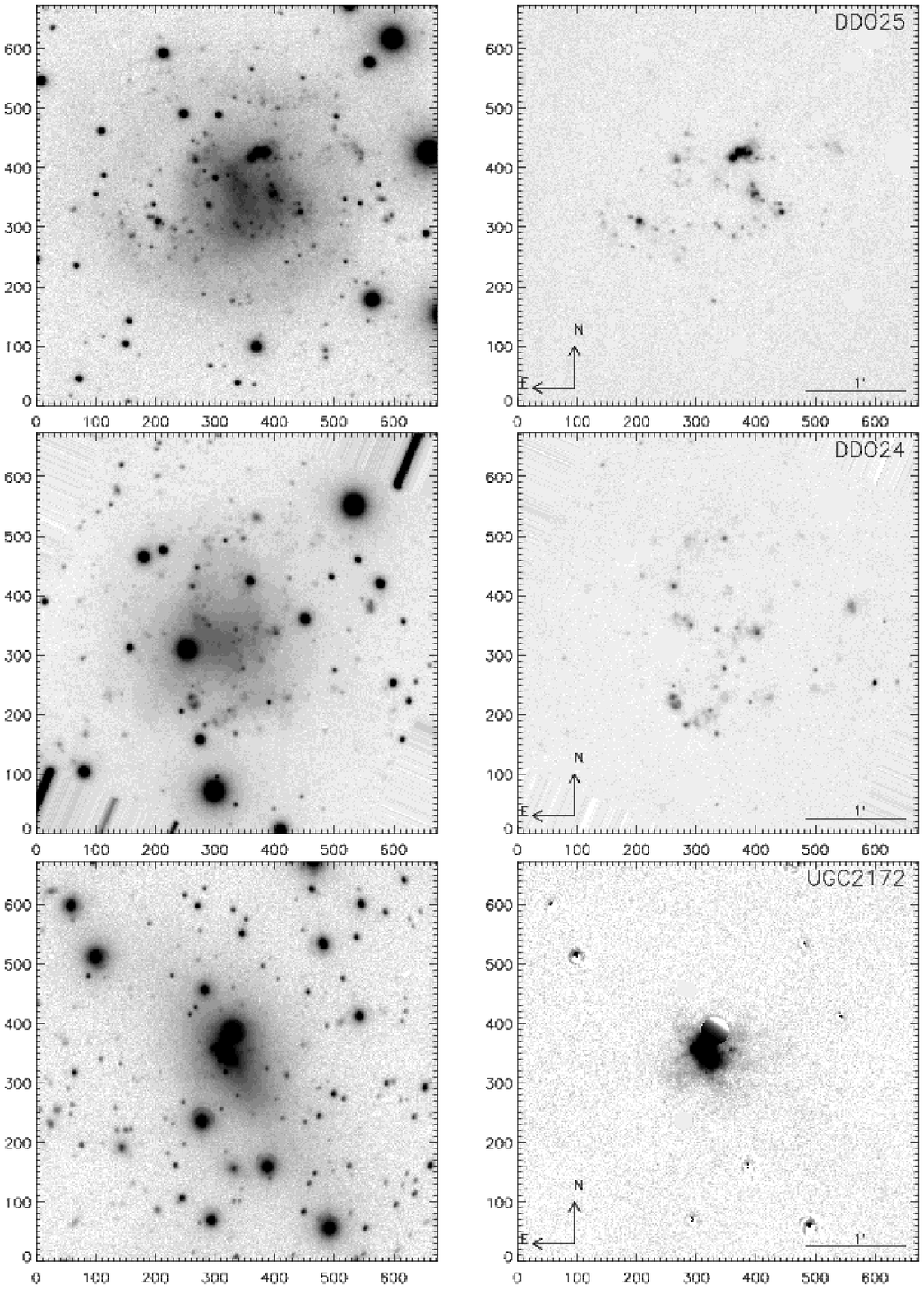}

\includegraphics[scale=0.8]{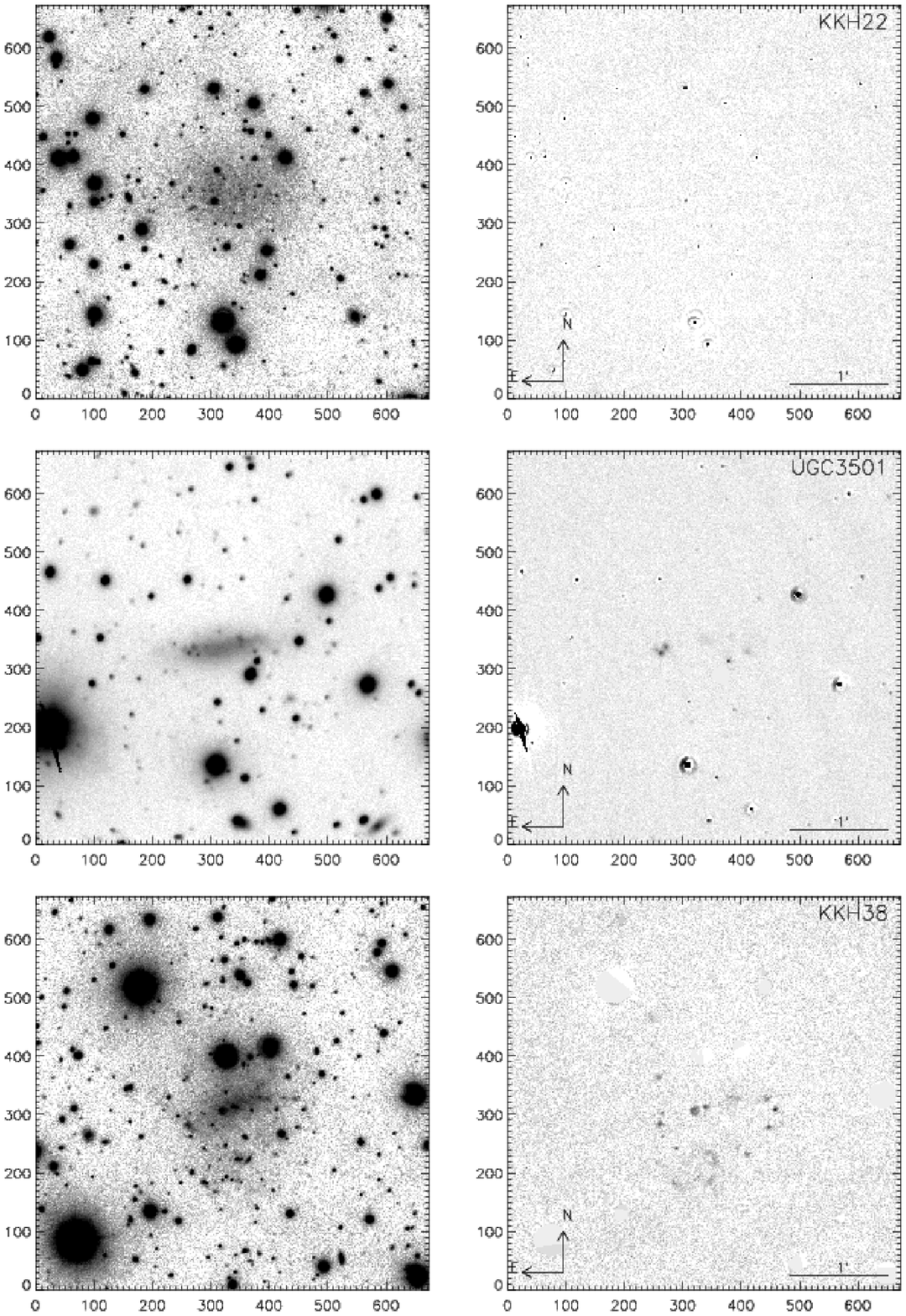}

\includegraphics[scale=0.8]{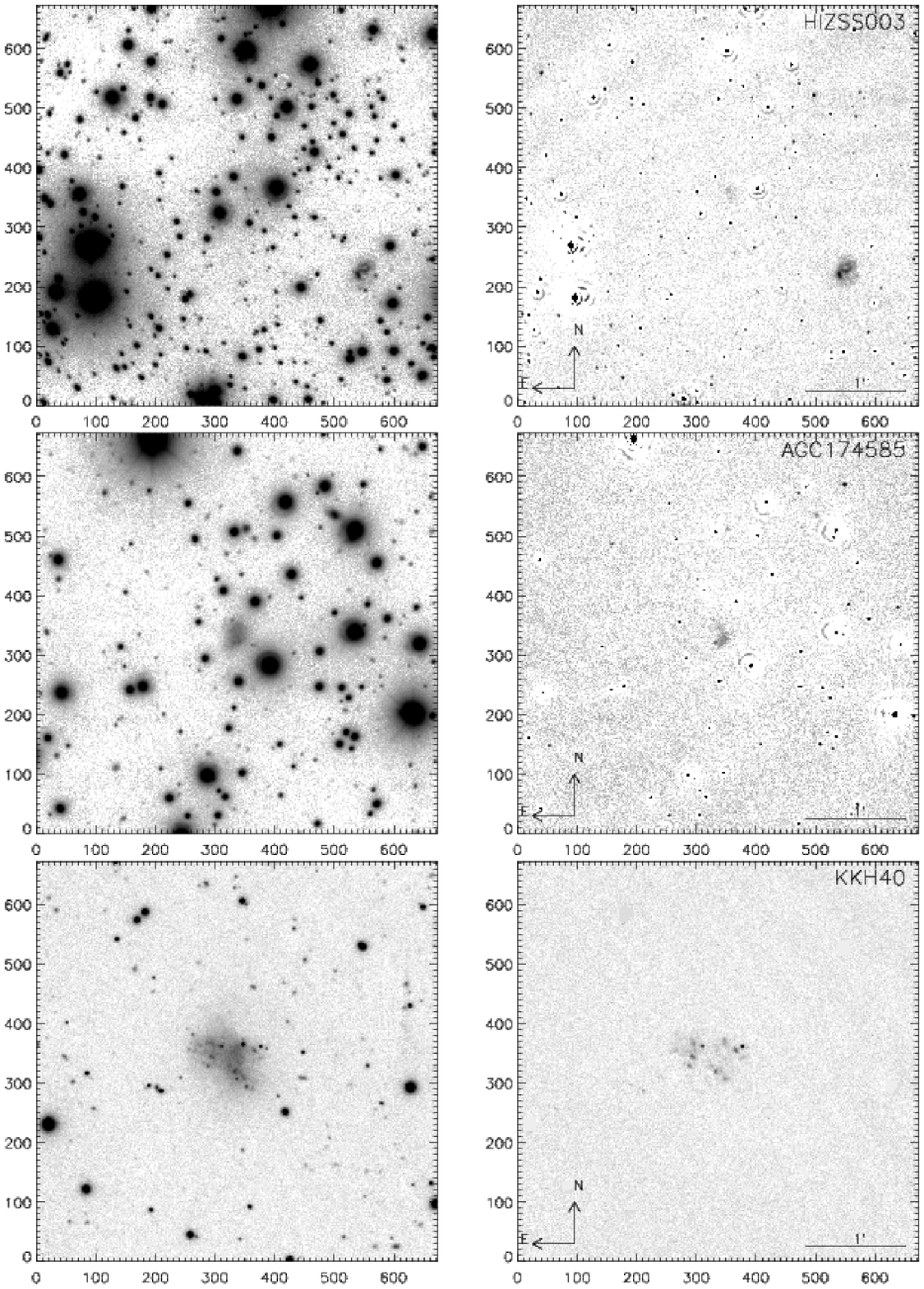}

\includegraphics[scale=0.8]{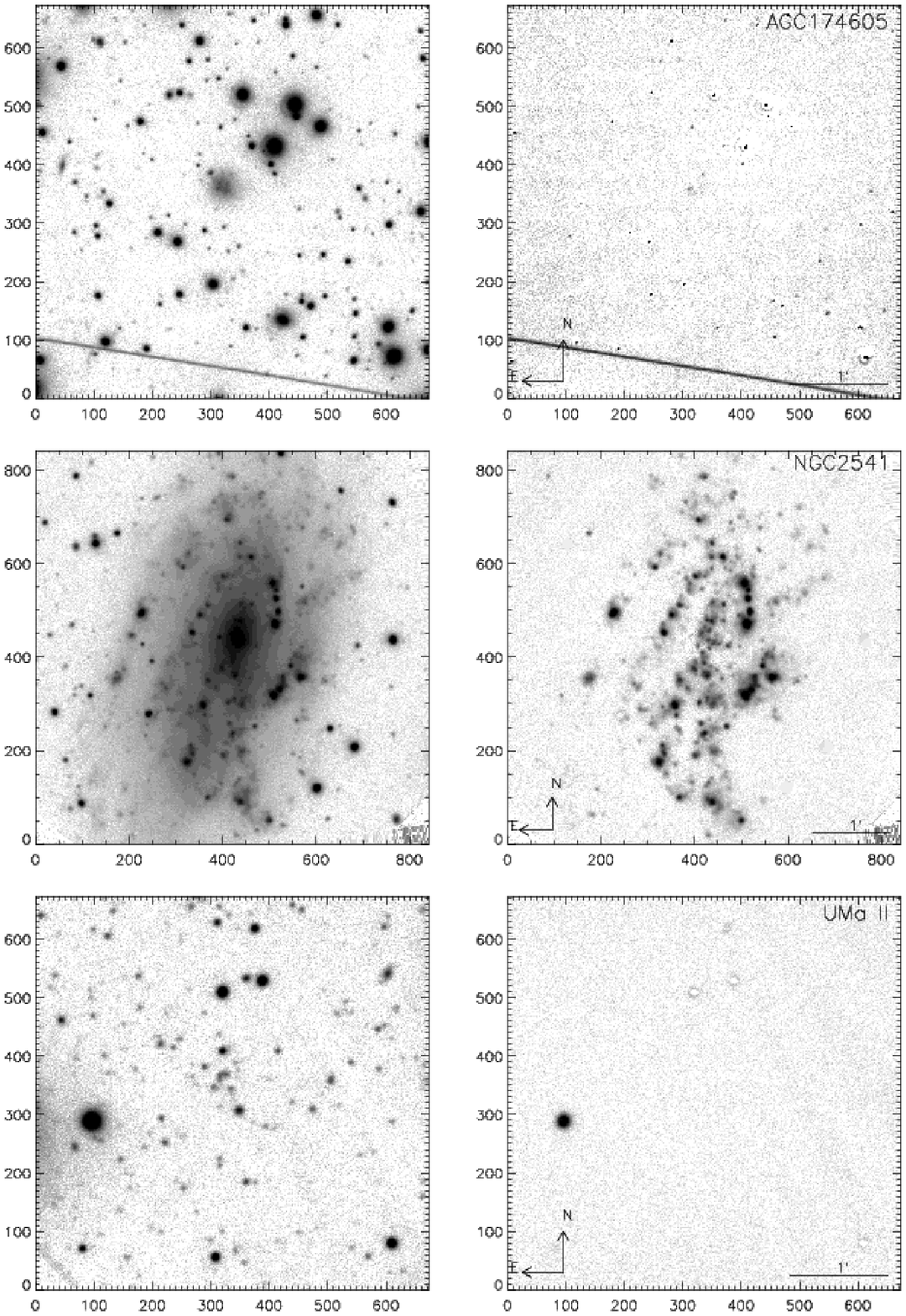}

\includegraphics[scale=0.8]{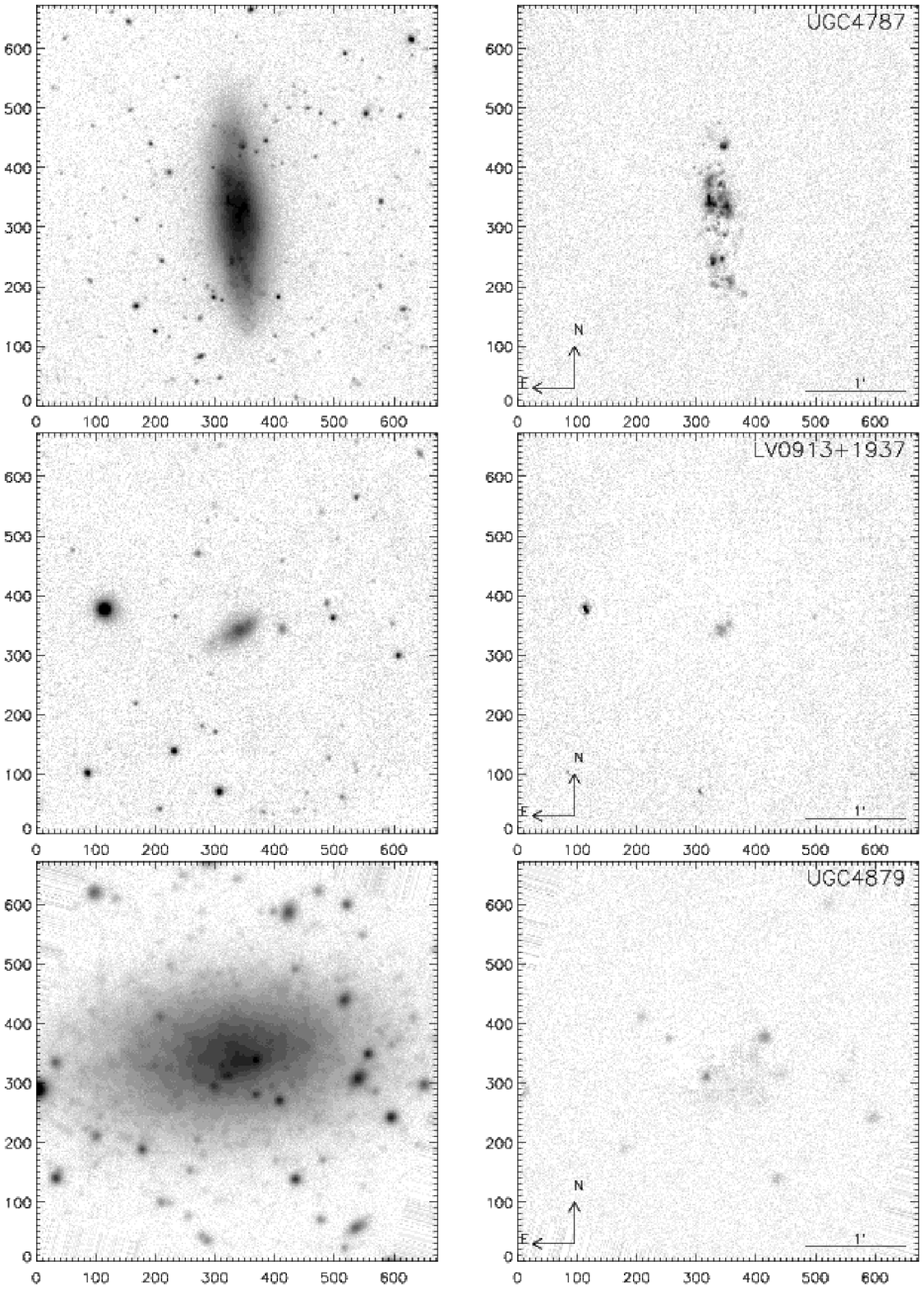}

\includegraphics[scale=0.8]{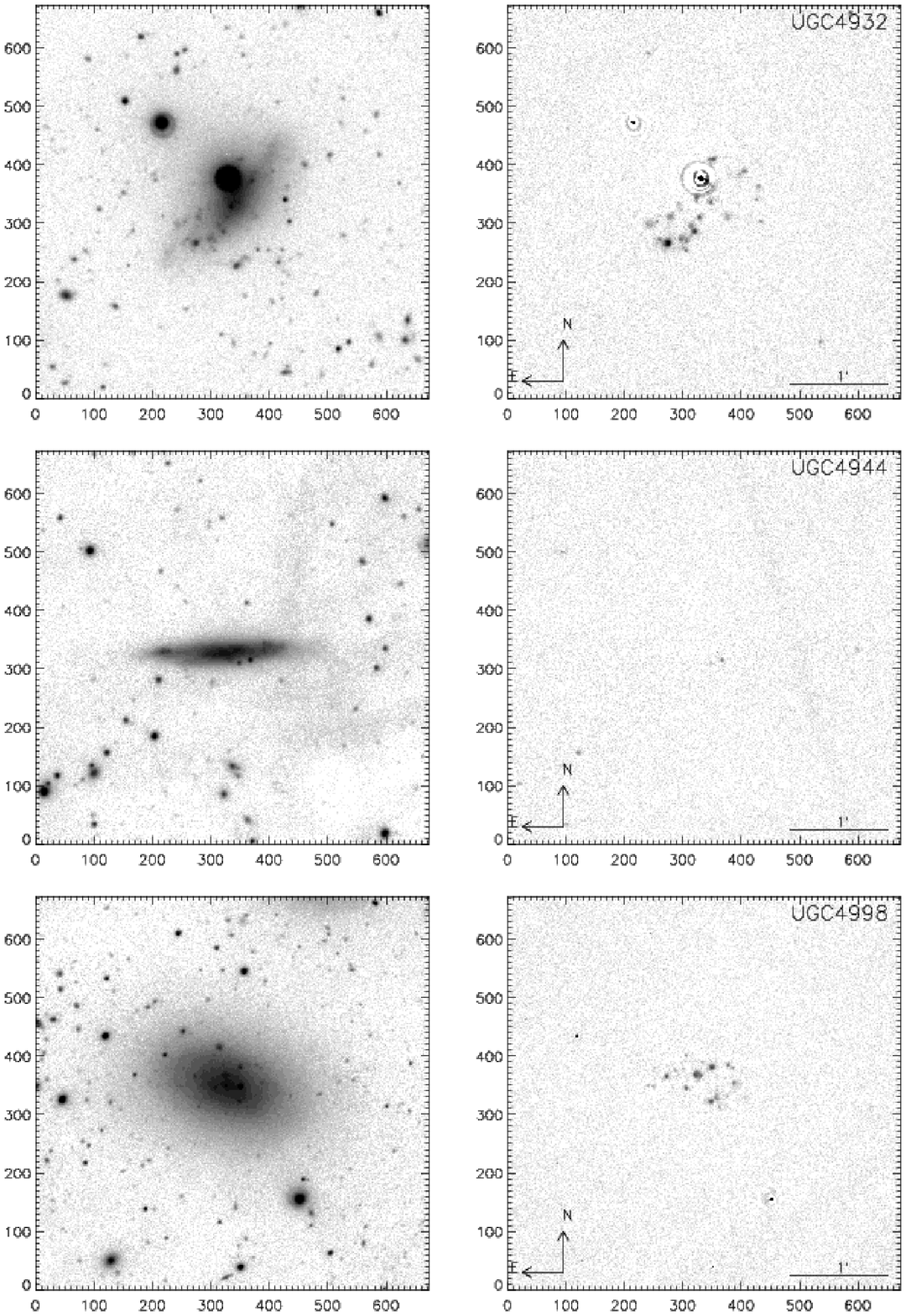}

\includegraphics[scale=0.8]{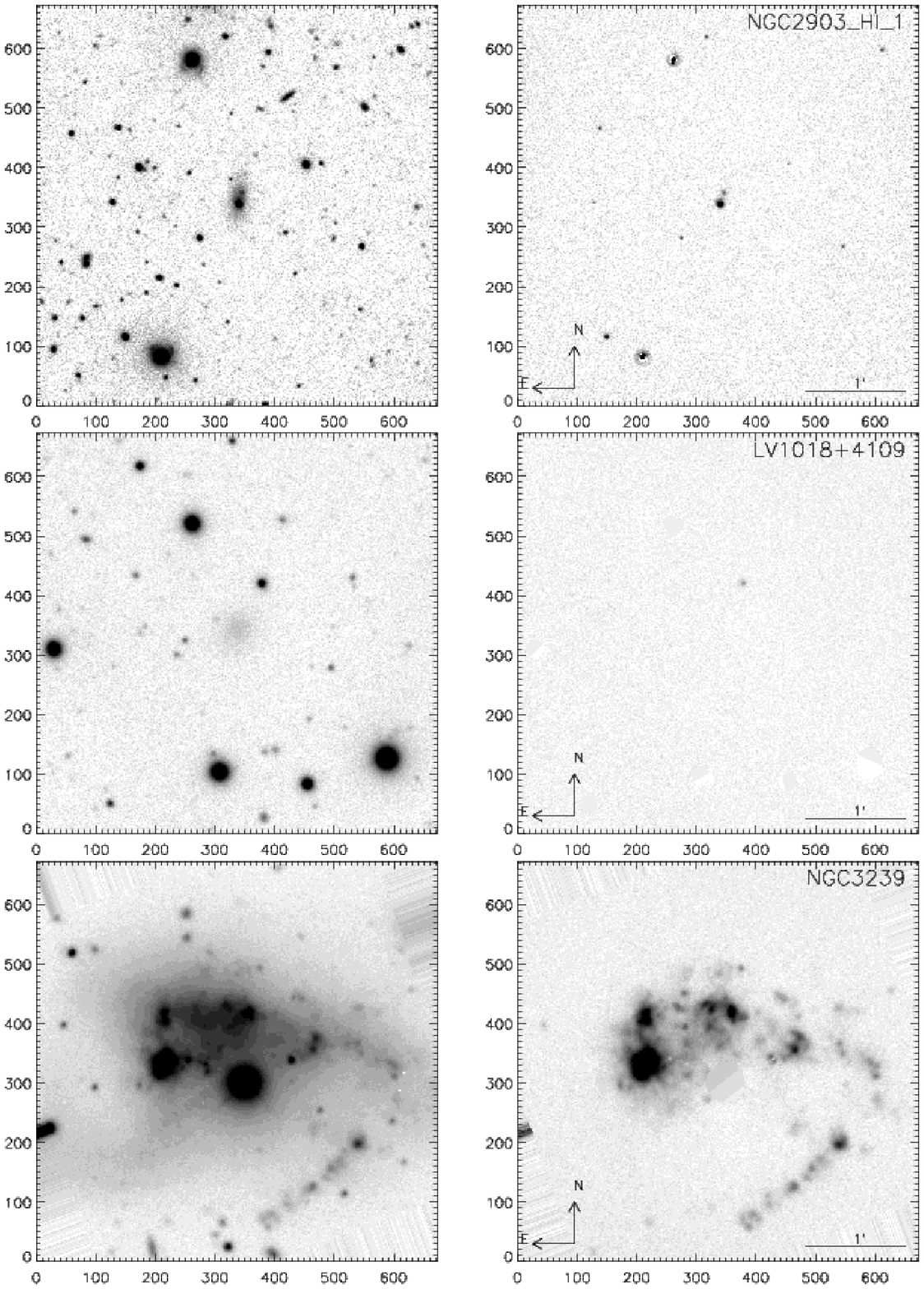}

\includegraphics[scale=0.8]{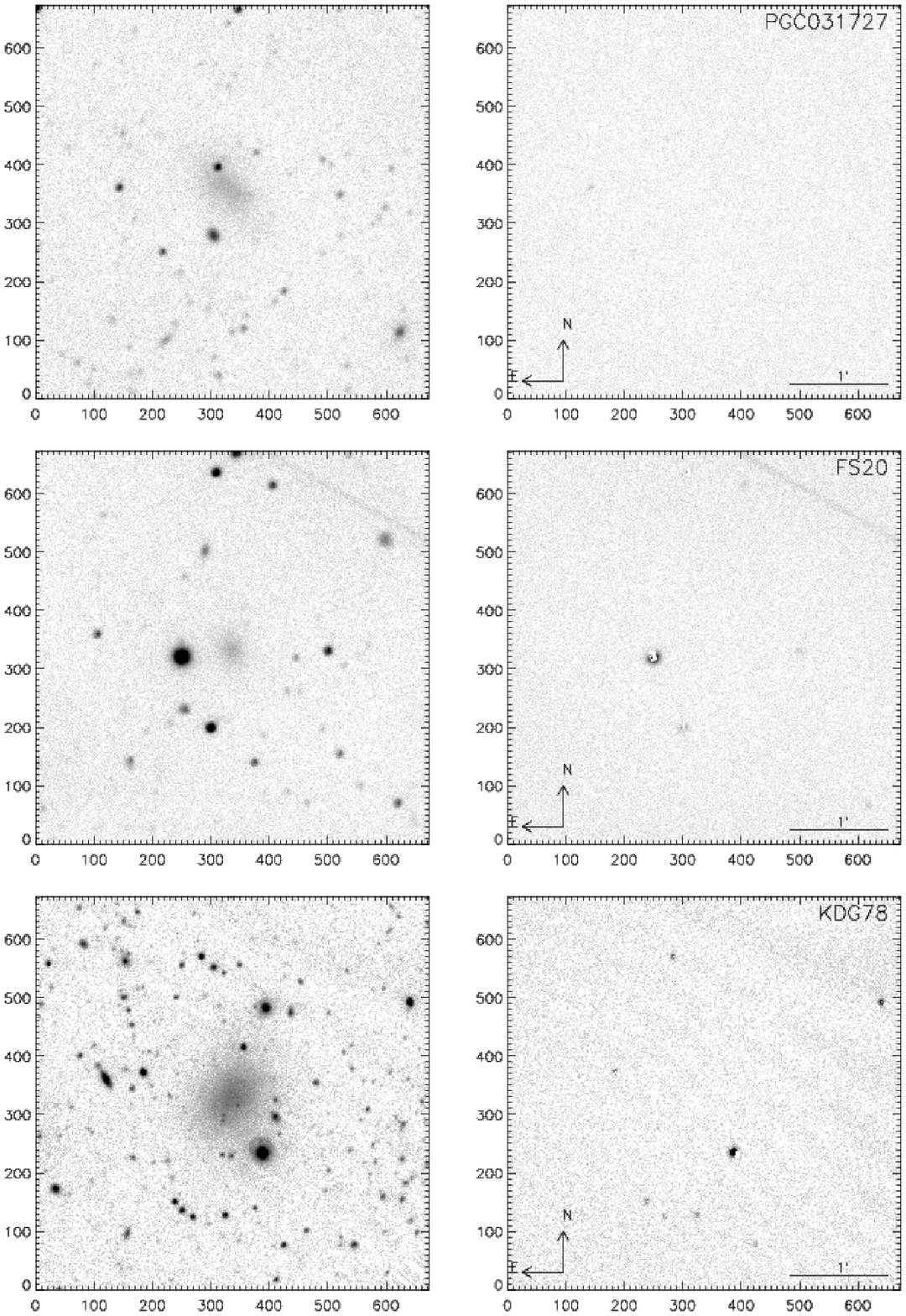}

\includegraphics[scale=0.8]{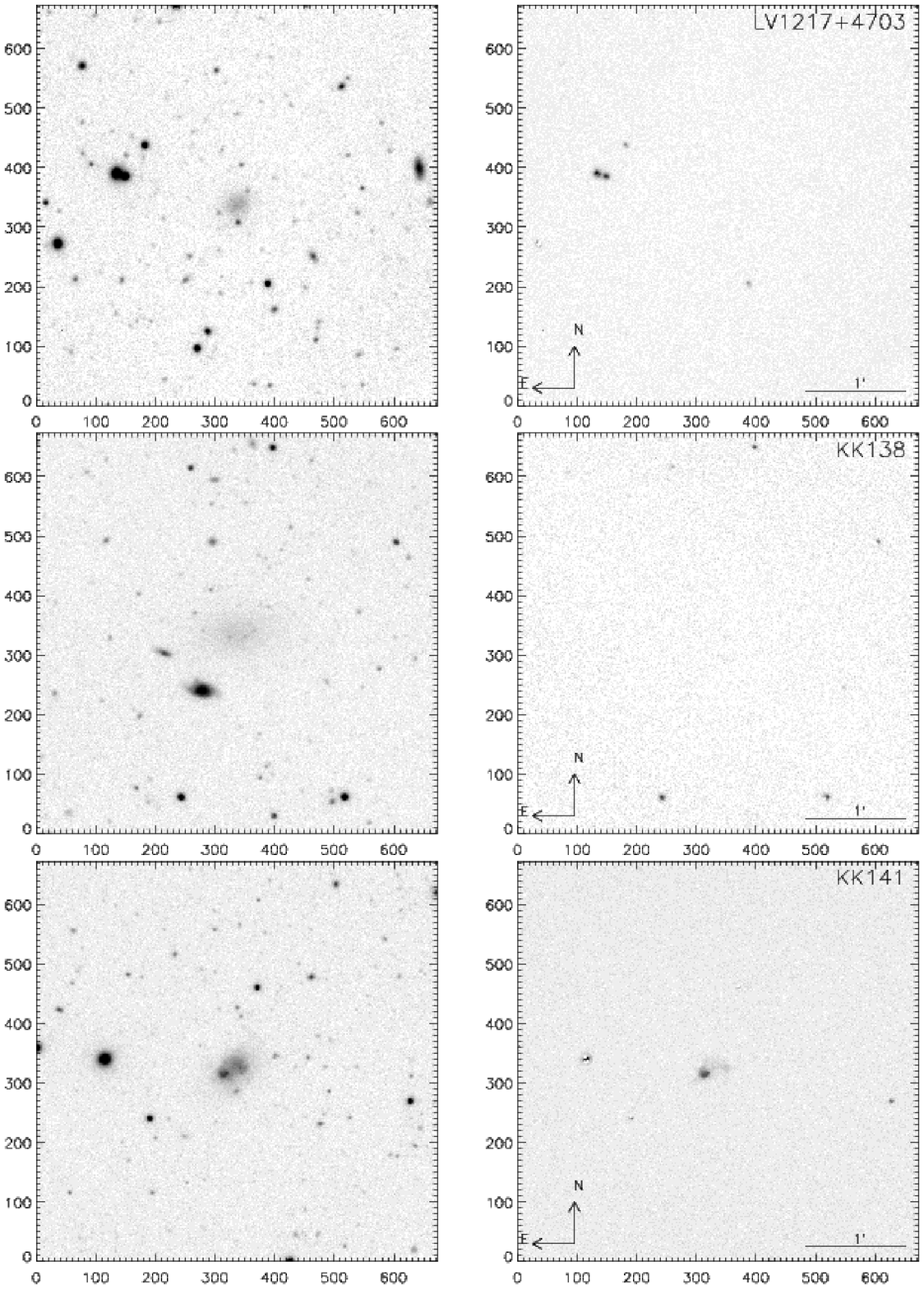}

\includegraphics[scale=0.8]{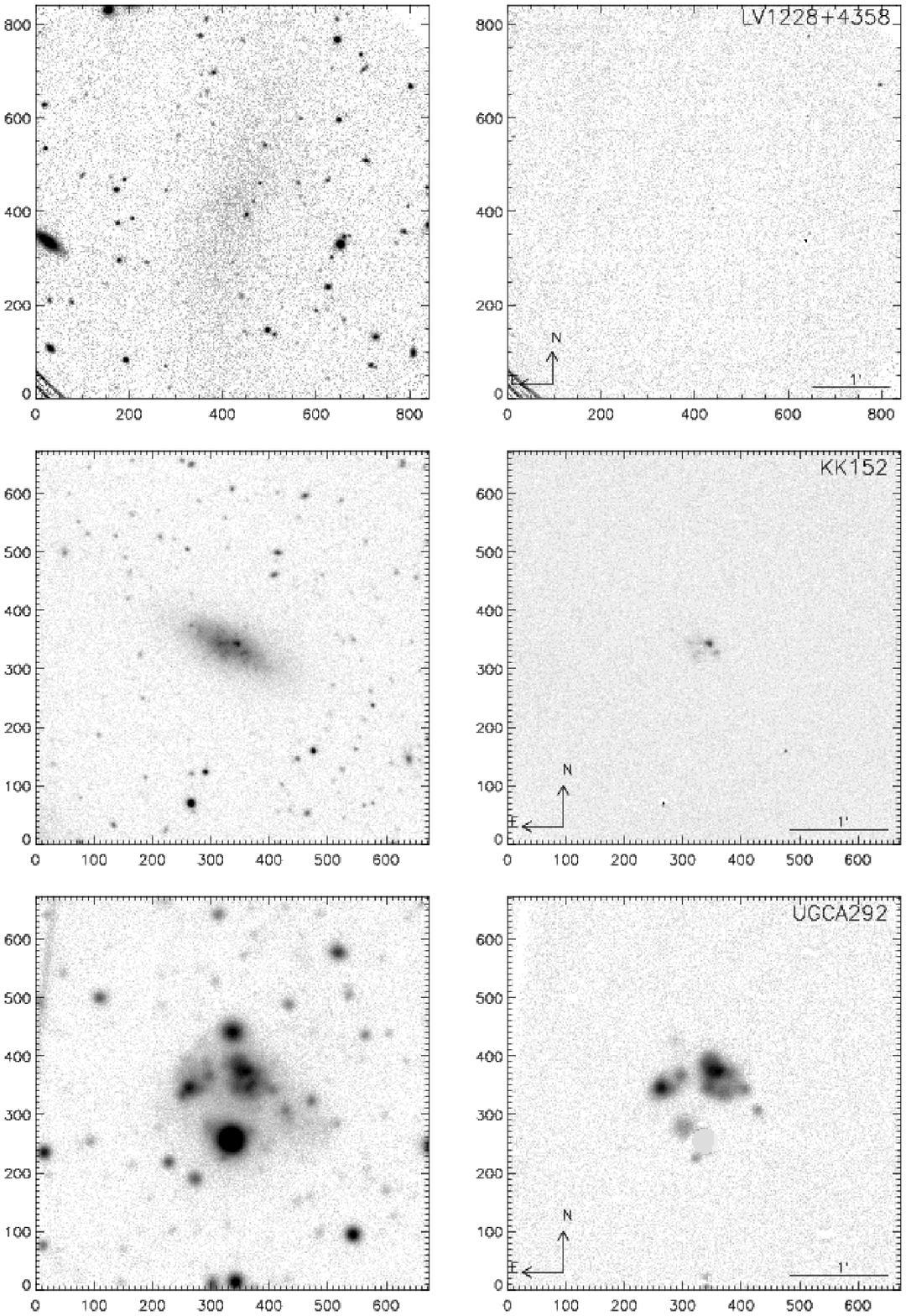}

\includegraphics[scale=0.8]{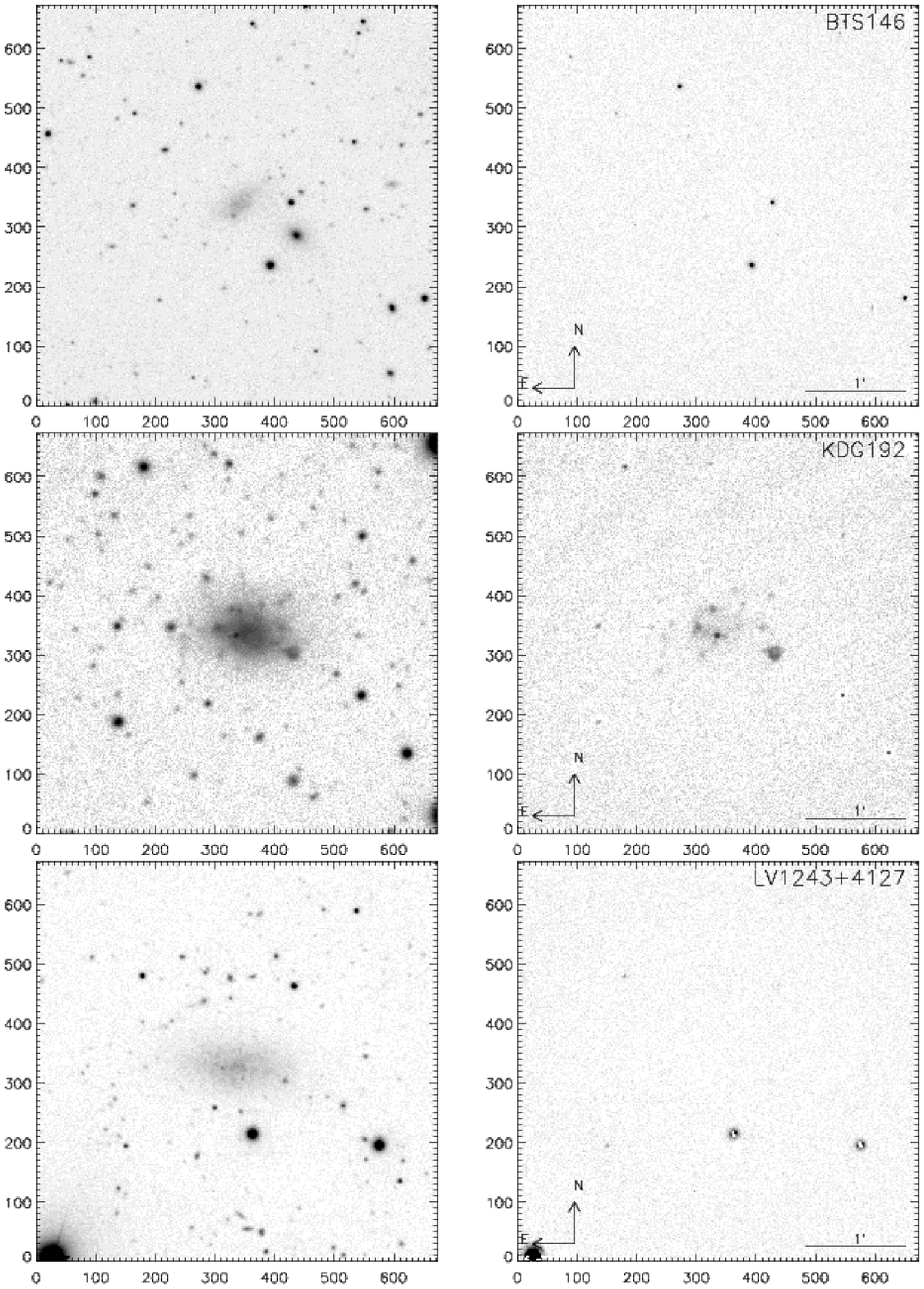}

\includegraphics[scale=0.8]{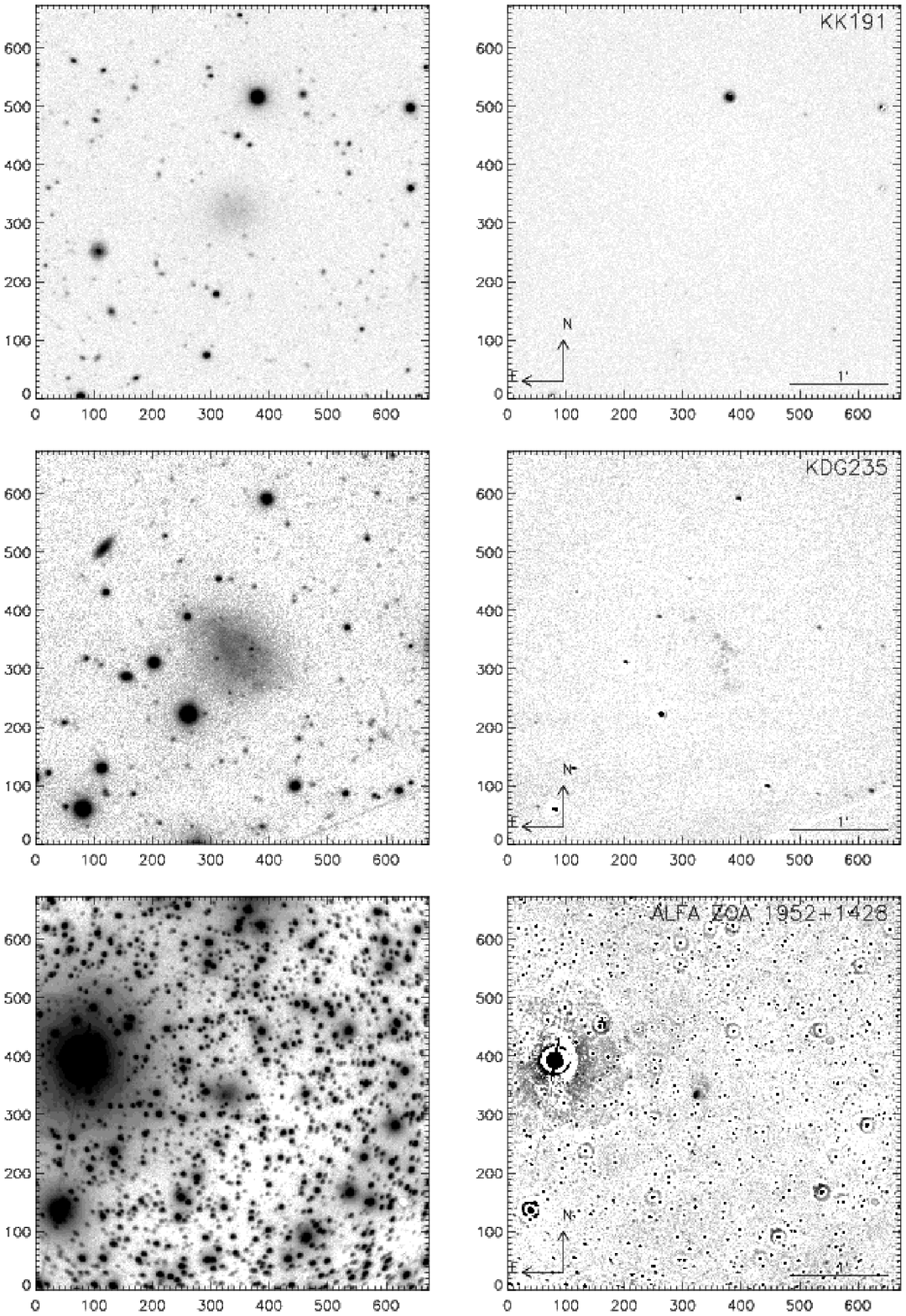}

\includegraphics[scale=0.8]{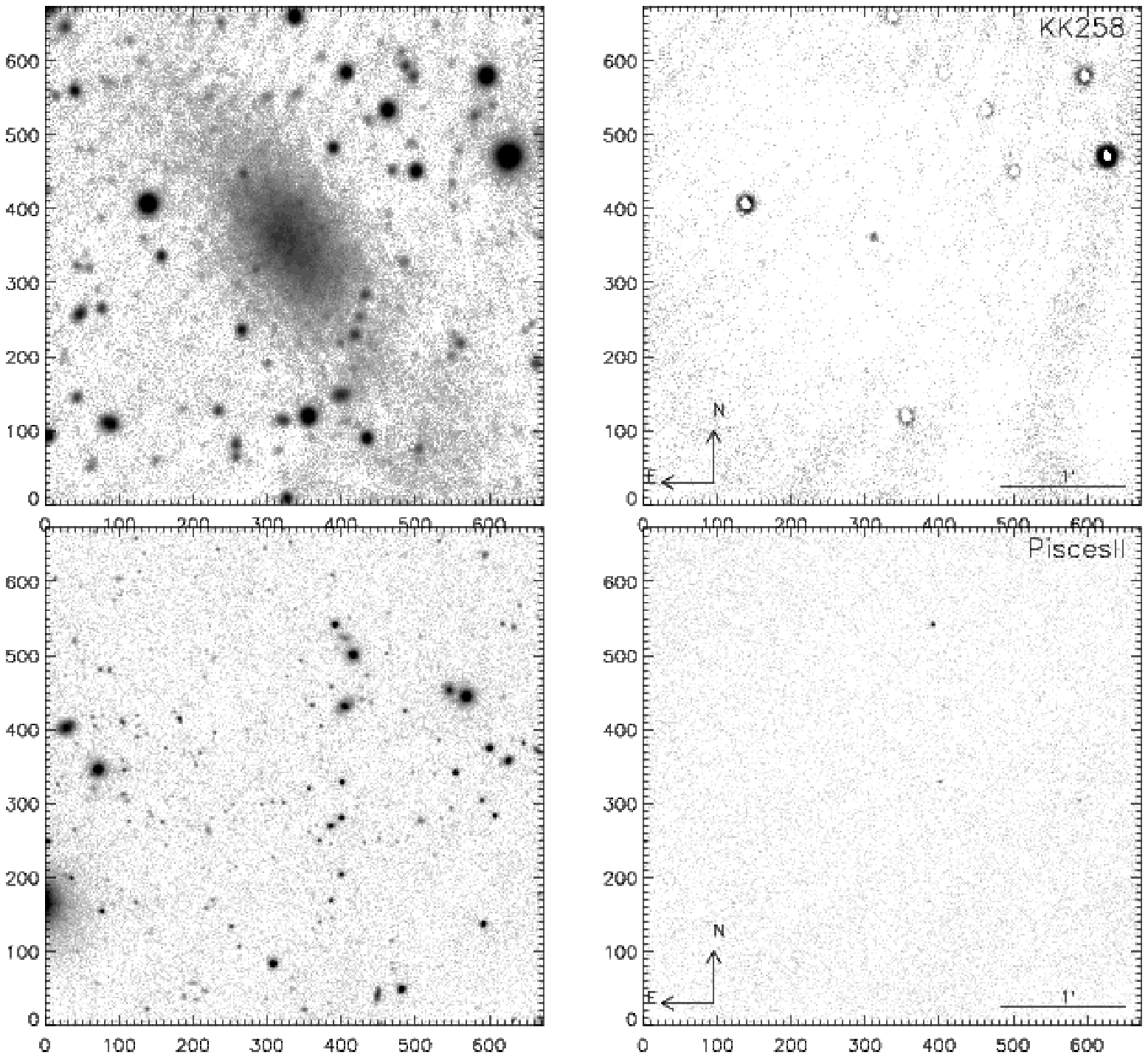}

%\endinput

\begin{thebibliography}{99}
 %1
\bibitem{kar2005:Kaisin_n}
 I.~D.~Karachentsev, S.~S.~Kaisin, Z.~Tsvetanov, and H.~Ford,  \aaa {\bf 434}, 935 (2005).
 %2
\bibitem{kai2006:Kaisin_n}
 S.~S.~Kaisin and I.~D.~Karachentsev, Astrophysics {\bf 49}, 287 (2006).

 %3
\bibitem{kar2007:Kaisin_n}
 I.~D.~Karachentsev and S.~S.~Kaisin,  \aj {\bf 133}, 1883 (2007).

%4
\bibitem{kai2008:Kaisin_n}
S.~S.~Kaisin and I.~D.~Karachentsev,  \aaa {\bf 479}, 603 (2008).

%5
\bibitem{kar2010:Kaisin_n}
I.~D.~Karachentsev and S.~S.~Kaisin,  \aj {\bf 140}, 1241 (2010).

%6
\bibitem{kai2011:Kaisin_n}
 S.~S.~Kaisin, I.~D.~Karachentsev, and E.~I.~Kaisina,  Astrophysics {\bf 54}, 315 (2011).

 %7
 \bibitem{kai2013:Kaisin_n}
 S.~S.~Kaisin and I.~D.~Karachentsev, Astrophysics {\bf 56}, 305 (2013).

%8
\bibitem{kar2013:Kaisin_n}
I.~D.~Karachentsev, D.~I.~ Makarov, and E.~I.~Kaisina,  \aj {\bf
145}, 101 (2013).

%9
\bibitem{kai2012:Kaisin_n}
E.~I.~Kaisina, D.~I.~Makarov, I.~D.~Karachentsev, and
S.~S.~Kaisin, \ab {\bf 67}, 115, (2012).

%10
\bibitem{afa2005:Kaisin_n}
V.~L.~~Afanasiev, A.~V.~Moiseev, Astronomy Letters {\bf 31}, 194, (2005).

%11
\bibitem{oke1990:Kaisin_n}
J.~B.~Oke,  \aj {\bf 99}, 1621 (1990).

%12
\bibitem{schle1998:Kaisin_n}
 D.~J.~Schlegel, D.~P.~Finkbeiner, and M.~Davis,  \apj {\bf 500}, 525 (1998).

 %13
 \bibitem{ken1998:Kaisin_n}
 R.~C.~Kennicutt,  \araa {\bf 36}, 189 (1998).

%14
\bibitem{ver2001:Kaisin_n}
 M.~A.~W.~Verheijen, \apj {\bf 563}, 694 (2001).

%15
\bibitem{lee2009:Kaisin_n}
 J.~C.~Lee, R.~C.~Kennicutt, J.~G.~Funes, et al., \apj, {\bf 692}, 1305 (2009).

%16
\bibitem{ken2008:Kaisin_n}
 R.~C.~Kennicutt, J.~C.~Lee, J.~G.~Funes, et al., \apjs {\bf 178}, 247 (2008).

%17
\bibitem{Kar2013:Kaisin_n}
 I.~D.~Karachentsev and  E.~I.~Kaisina, \aj {\bf 146}, 46 (2013).

%18
\bibitem{gil2007:Kaisin_n}
 A.~Gil de Paz, S.~Boissier, B.~F.~Madore, et al., \apjs {\bf 173}, 185 (2007).

 %19
 \bibitem{sil2005:Kaisin_n}
 D.~R.~Silva, P.~Massey, K.~DeGioia-Eastwood, and P.~A.~Henning, \apj {\bf 623}, 148 (2005).

%20
\bibitem{beg2005:Kaisin_n}
 A.~Begum, J.~N.~Chengalur, I.~D.~Karachentsev, and M.~E.~Sharina, \mnras {\bf 359}, L53 (2005).

%21
\bibitem{mas2003:Kaisin_n}
  P.~Massey, P.~A.~Henning, and R.~C.~Kraan-Korteweg, \aj {\bf 126}, 2362 (2003).

 %22
 \bibitem{zuk2006:Kaisin_n}
 D.~B.~Zucker, V.~Belokurov, N.~W.~Evans, et al., \apj {\bf 650}, L41 (2006).

%23
\bibitem{irw2009:Kaisin_n}
J.~A.~Irwin, G.~L.~Hoffman, K.~Spekkens, et al., \apj {\bf 692},
1447 (2009).

%24
\bibitem{Kar2007:Kaisin_n}
I.~D.~Karachentsev, V.~E.~Karachentseva, and W.~K.~Huchtmeier,
Astron. Lett. {\bf 33}, 512 (2007).

%25
\bibitem{mar2012:Kaisin_n}
 D.~Martinez-Delgado, A.~J.~Romanowsky, R.~J.~Gabani, et al., \apj {\bf 748}, L24 (2012).

%26
\bibitem{van2000:Kaisin_n}
L.~van~Zee, \apj {\bf 543}, L31 (2000).

%27
\bibitem{mci2011:Kaisin_n}
T.~McIntyre, R.~F.~Minchin, E.~Momjian, et al., \apj {\bf 739},
L26 (2011).

%28
\bibitem{bel2010:Kaisin_n}
V.~Belokurov, M.~G.~Walker, N.~W.~Evans, et al., \apj {\bf 712},
L103 (2010).

%29
\bibitem{pfl2007:Kaisin_n}
J.~Pflamm-Altenburg, C.~Weidner, and P.~Kroupa,  \apj {\bf 671},
1550 (2007).
\end{thebibliography}
\end{document}